**Title:** Plasmonic Color Filters Enable Label-Free Plasmon-Enhanced Array Tomography with sub-diffraction limited resolution

*Kristian Caracciolo[1,2,4], Eugeniu Balaur[1,4], Erinna F. Lee[3,4,5,6], W. Douglas Fairlie[3,4,5,6], Julian Ratcliffe[2], Eric Hanssen[7], David Hoxley[1,4], Jacqueline Orian[3,4], Chad Johnson[2], Lu Yu[8], Kang Han[8], Wei Xiang[8], Brian Abbey[1,4]*

1. Department of Mathematical and Physical Sciences, School of Computing, Engineering and Mathematical Sciences, Bundoora, Victoria, Australia
2. La Trobe University Bioimaging Platform, Bundoora, Victoria, Australia
3. Department of Biochemistry and Chemistry, School of Agriculture, Biomedicine and Environment, Bundoora, Victoria, Australia
4. La Trobe Institute for Molecular Science, La Trobe University, Bundoora, Victoria, Australia
5. Olivia Newton John Cancer Research Institute, 145 Studley Rd, Heidelberg, Victoria, Australia
6. School of Cancer Medicine, Bundoora, Victoria, Australia
7. Ian Holmes Imaging Centre and Department of Biochemistry and Pharmacology and ARC Centre for Cryo Electron Microscopy of Membrane Proteins, Bio21 Molecular Science and Biotechnology Institute, University of Melbourne, Parkville, Victoria, Australia
8. School of Computing, Engineering and Mathematical Sciences, La Trobe University, Melbourne, 3086, Australia

Correspondence to: Eugeniu Balaur e.balaur@latrobe.edu.au and Brian Abbey b.abbey@latrobe.edu.au

Three-dimensional (3D) imaging of the subcellular organisation and morphology of cells and tissues is essential for understanding biological function. Although staining is the most widely used approach for visualising biological samples under a microscope, the intracellular refractive index (RI) has been proposed as a potential biophysical marker that could supplement or even surpass the sensitivity of current histological methods. Hence, the development of new, highly sensitive, label-free techniques that can detect changes in the intracellular RI is extremely desirable for biomedical imaging. The recent development of plasmonic metamaterials, designed to mimic a traditional microscope slide, have made it possible to translate subtle changes in refractive index directly into color. This approach enables label-free visualization of tissue microstructure using standard histological slide preparation methods. Here we demonstrate ultramicrotome-assisted optical plasmon-enhanced (PE) array tomography and correlate this with electron array tomography of the same sample embedded in resin. The approach enables axially super-resolved label-free imaging of whole cells in the range of 30−200 nm and shows great potential for multimodal three-dimensional colorimetric histology at the (sub-) organelle level.

## 1. Introduction

The functional behaviour of cells and tissues is inherently related to their structure and morphology, which varies on lengthscales ranging from nanometers up to whole cell structures. As such, new imaging techniques with high spatial resolution that are capable of resolving these structures in three dimensions (3D) are highly desirable. One method of achieving super-resolution imaging in 3D is to combine electron microscopy (EM) with serial microtomy in a technique known as array tomography[1]. This technique exploits the high sensitivity of electron microscopy in order to generate images from ultrafine tissue sections with individual thicknesses below the Abbe diffraction limit. By serially imaging the sections and combining them into a 3D volume, a tomographic image of the sample can be generated[2].

Although EM tomography can achieve nanometer-scale imaging, EM is limited to grayscale images in which the signal intensity at each pixel corresponds to a single number that represents the proportional number of electrons either transmitted through a thin sample or emitted from a surface. In order to supplement EM a variety of optical microscopy techniques exist, many of which are based on fluorescence microscopy, which provides molecular specificity and generates signals that can be localised to the emitter with a spatial resolution at the nanometer scale[3]. However, in situations where an appropriate molecular biomarker is unavailable, or photobleaching poses a problem, or when complementary optical morphological or chemical information is required, the utilization of alternate microscopy techniques becomes necessary. This has motivated the development of the field of 'label-free' biological imaging which exploits the inherent specific interactions of cells and tissues with light in order to render their chemistry and morphology visible. One of the most well-developed techniques for label-free imaging is phase contrast microscopy. Phase contrast microscopy is sensitive to the refractive index (RI) of cells and tissues and is able to render features visible which are not apparent in absorption contrast alone. Techniques which exploit the RI of biological samples in order to measure their morphology via changes in optical density have been hugely successful in the field of bioimaging and have been used extensively for understanding and diagnosing disease[4]. However, when examining the ultrafine histological sections used in array tomography, which are in the range of 30-200 nm in thickness[1], even phase contrast techniques typically have insufficient sensitivity to detect intracellular structural variations in biological samples[5, 6]. In addition, similar to EM, the images generated are typically rendered in grayscale, and are only colored artificially after collection.

The recent development of plasmon-enahanced biosensing and imaging techniques that utilize visible light and exploit the extreme sensitivity of surface plasmons (SPs) to changes in the near-surface RI offers a potential solution to these problems[7, 8]. Surface plasmons are coherent oscillations of electrons that under suitable excitation conditions, occur at the metal-dielectric interface. The electromagnetic wave encompassing both the charge motion of the electrons (SPs) and the associated electric field extending into both the metal and dielectric layers is known as a Surface Plasmon Polariton (SPP). Planar optical metasurfaces consisting of periodic arrays of subwavelength electromagnetic structures generating both SPs and localised surface plasmons (LSPs) have been widely employed as optical color filters[9-11]. A schematic of the multilayer plasmonic color filters used for the present study is shown in Fig. 1a; a corresponding, top-down, example Scanning Electron Microscopy (SEM) image of the device surface is shown in Fig. 1b. The devices have the same overall dimensions as conventional microscope slides (i.e. 75 × 25 × 1 mm) and are encased in a chemically inert layer to ensure that they do not degrade over time and are biocompatible. Because of the specific structure of the plasmonic device, incident light in transmission excites SPs everywhere it impinges due to momentum coupling between the incident photons and the free electrons. The oscillating electrons generate an evanescently decaying electromagnetic field normal to the surface of the nanoapertures (Fig. 1c) extending into the sample (Fig. 1c inset). The output of these plasmonic color filters can be readily tuned by tailoring the device geometry[12, 13], incident polarisation[14-16], or even temperature[17] in order to modify the amplitude and phase of the transmitted or reflected light. The resulting position of the resonant transmission peaks associated with the plasmonic device vary depending on the local sample RI producing a range of different characteristic colors (Fig. 1d).

Tomography of biological cells and tissues employing plasmonically active nanoparticles has previously been demonstrated, exploiting their electronic properties to enhance certain local characteristics of the sample such as optical scattering or diffusion. This is typically achieved by integrating plasmonic nanoparticles into other optical tomographic imaging techniques such as optical coherence tomography[18, 19] or photoacoustic tomography[20]. The individual nanoparticles effectively act as a biomarker analogous to conventional fluorescence imaging providing a window into the local optical properties of the sample[21]. In contrast to the label-free microscopy used here, images are generated by locating the individual nanoparticles and determining how they are distributed and bind to the cells or individual molecules.

## Fig 1: Device structure and principles of SPP generation

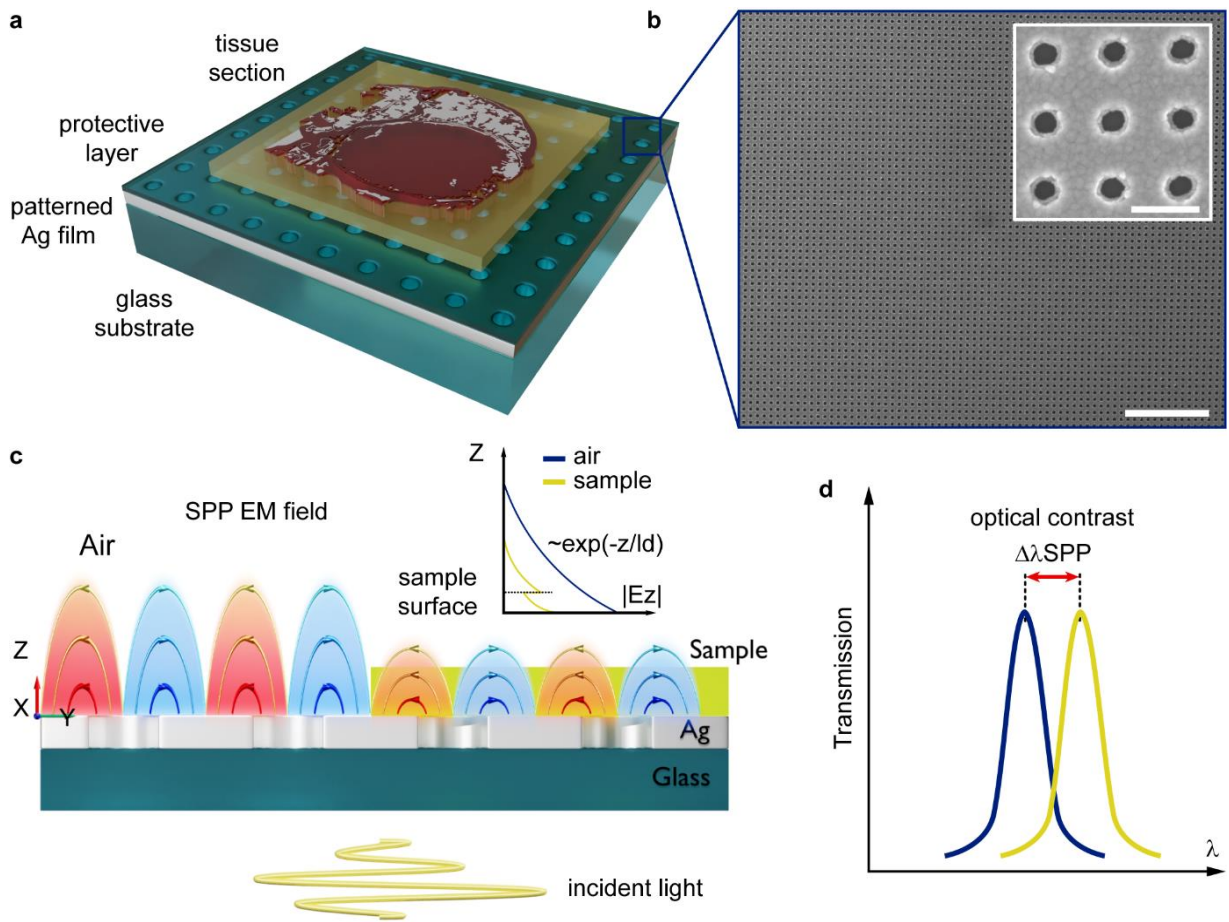

a) Schematic diagram of multilayer plasmonic device structure. b) Top-down scanning electron microscopy (SEM) image of the surface of the device. The scale bar in the main image is 5 μm. The scale bar in the enlarged corner image is 400 nm. c) Diagram showing evanescent electromagnetic field in air and sample generated by SPPs extending outward in a direction normal to the surface. Inset shows exponential decay of the magnitude of the *Z* component of the electric field, |Ez| , as a function of distance from the plasmonic device surface. The decay behaviour both with (yellow curve) and without (blue curve) a sample in place is shown. d) The wavelength associated with the resonant transmission peaks is extremely sensitive to the local refractive index (RI), illustrated schematically by the two curves. The curves represent one of the resonance transmission peaks with (blue) and without (yellow) a sample in place.

Here we demonstrate the first integration of planar plasmonic color filters with array tomography (30-150 nm axial resolution), enabling 3D colorimetric histology of biological tissues and cells. Not only does PE array tomography provide enhanced optical contrast, for uniform thickness specimens the colors can be assumed to vary with the sample RI providing a window into the local chemical composition. The use of plasmonic color filters for chemical and biological sensing on the basis of RI has been developed extensively over the past decade[22], but it is only very recently that these types of devices have been exploited for plasmon-enhanced colorimetric histology[6].

Our approach for correlative tomographic plasmon-enhanced/SEM bioimaging allows us to use standard electron array tomography sample preparation and procedures in conjunction with conventional optical brightfield imaging without the need for any additional equipment or specialised setups. As a consequence of the nanofabrication (see Methods) the surface is conducting enabling direct correlative imaging via electron array tomography. This provides multiple complementary contrast mechanisms for the exact same intracellular structures by stacking and aligning serial ultrafine tissue sections to perform the first PE array tomography (Fig. 2a). This demonstration opens the way to superresolution PE array tomography of whole cells and tissues correlated with EM using standard laboratory workflows. The three-dimensional colorimetric images generated provide a comprehensive view of how specific biological components vary across the histological block volume.

## 2. Results

Contrast Enhancement

Contrast enhancement in plasmon-enhanced colorimetric histology is linked to the phenomenon of Extraordinary Optical Transmission (EOT) which occurs due to SPPs[23] generated via our plasmonically active microscope slides[24]. Silver and gold are routinely used to produce SPPs due to their high conductivity and specific electronic properties in the optical regime although devices using alternate materials such as aluminum[25] and all-dielectric nanophotonics have also been successfully demonstrated[25-27]. The wavevector of the SPPs is highly sensitive to the periodicity of the pattern in the film, and the RI of the sample which sits within the evanescent field generated by the SPs. This dictates the resonant peak structure, which is a feature of the optical transmission for these devices resulting in a strong color contrast that spans the entire visible spectrum and is sensitive to the local sample RI. In addition, when the sample is thinner than the typical decay length of the evanescent electric field generated by the SPs (~ 200 nm), the color output is also sensitive to changes in the sample thickness. The changes in both color and intensity due to an extreme sensitivity to changes in the local sample RI have been previously exploited to detect the attachment of biomolecules to surfaces[28, 29], monitor the presence of ion implantation-induced defects in thin films[30], and to differentiate cancer cells[6, 31].

In general, for ultrafine tissue sections with thicknesses well below the optical diffraction limit (typically ~300 nm), it is challenging for many label-free imaging techniques to generate significant contrast[5]. This is in part due to the fact that these techniques typically generate greyscale images which have limited dynamic range compared to color images. As an example, we have compared Differential Interference Contrast (DIC) imaging (Fig 2b) to colorimetric histology imaging (Fig. 2c) for a 120 nm thick section of mouse optic nerve tissue. An example lineout comparing the normalized intensity variation is shown in Fig. 2d. In the DIC image the edge of the tissue, where there is a strong gradient in the RI between the tissue and the surrounding air, is clearly visible but the internal morphological features are almost entirely absent. In the case of ultramicrotome tissue sections which are highly planar with smoothly varying structures, we generally observe only minor differences in contrast between phase-sensitive microscopy techniques and conventional brightfield imaging. In the case of DIC, even the observed edge contrast was found to disappear after coverslipping, due to the encapsulating mounting media having a similar RI to the tissue/resin. Conversely, we observed strong plasmon-enhanced colorimetric contrast for the ultramicrotome tissue sections, independent of the degree of edge contrast. This is related to the fact that plasmon-enhanced colorimetric histology is mediated by near-field electromagnetic interactions between the sample and plasmonic device. Here we observe that contrast in ultramicrotomed sections persisted, even down to section thicknesses of just 30 nm with no significant optically detectable variation in section thickness (Fig. S1).

**Fig 2: PE array tomography workflow and contrast enhancement.**

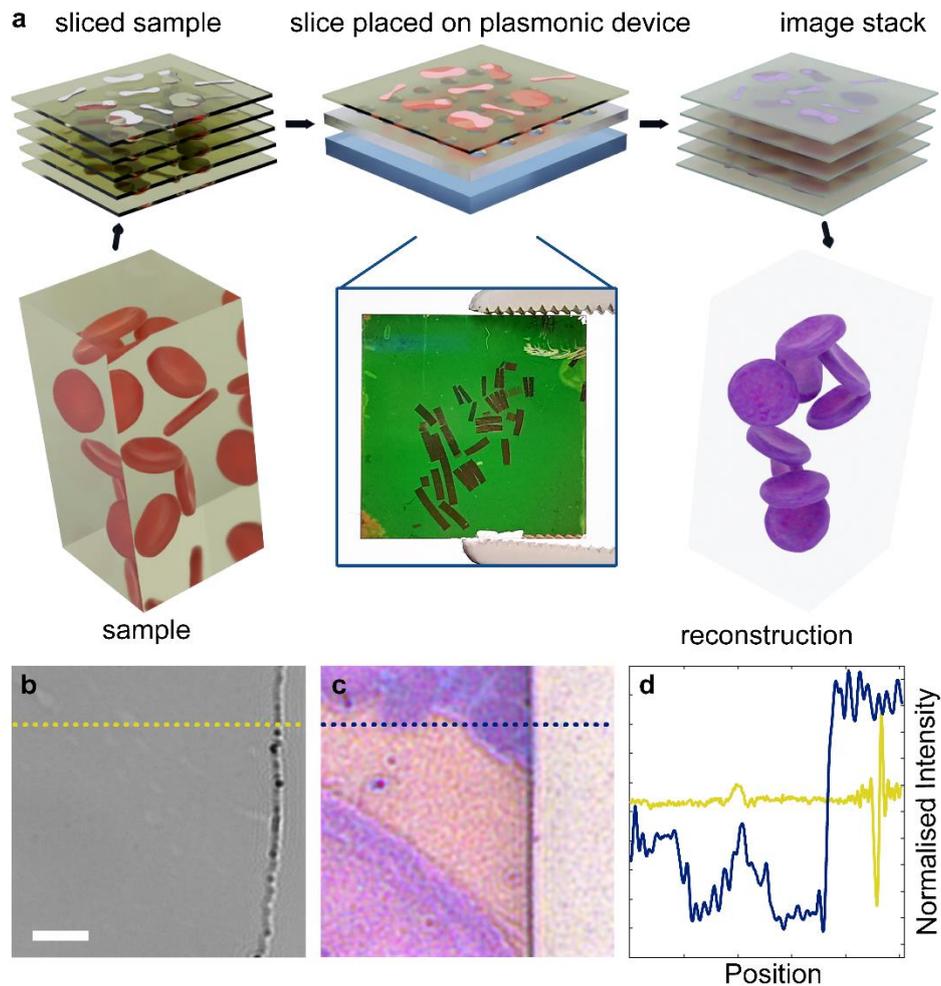

a) Schematic flowchart of PE array tomography protocol beginning with a whole resin block embedding e.g. cells or tissue. This is followed by ultramicrotomy and section stacking, slice alignment, and 3D reconstruction. Bottom middle – a photograph of a plasmonic device with resin sections placed on it. b) DIC microscopy image of a 120 nm thick unstained nerve tissue section on conventional glass microscope slide. Scalebar is 20 µm. c) Standard brightfield microscopy image of a 120 nm thick unstained nerve tissue section taken from the same block as b) placed on a plasmonically active microscope slide demonstrating contrast enhancement from plasmon-enhanced colorimetric histology. d) Normalised intensity lineout from the regions indicated by horizontal yellow and blue dashed lines in b) and c), respectively.

## Volume Reconstructions

PE array tomography was first performed using a well-characterised test sample composed of unstained polystyrene microspheres embedded in resin (Fig. 3a). According to the manufacturer's data these spheres had a diameter of between 4.8 μm to 5.8 μm. The PE array tomography volume images were formed from 115 and 96 individual ultramicrotome sections for the microsphere and cell samples respectively, each with a nominal thickness of 100 nm. The sections were optically imaged using a conventional brightfield microscope (see Methods) and stacked vertically along the Z-axis (axial direction) in order to create a 3D volume. The spheres had well-defined sharp edges which enabled an experimental determination of the lateral and axial spatial resolution (Fig. 3b, c). Here, the experimental spatial resolution values were found to be 301 nm and 153 nm in the lateral and axial planes, respectively i.e. the experimental axial resolution was ~ half the lateral resolution and substantially below the optical diffraction limit.

When the ultramicrotome section thickness is less than the evanescent decay length (~300 nm) variations in both thickness and RI will result in color changes on the plasmonic slide. When the section thickness is larger than the evanescent decay length (e.g. > 1 micron thick) then any thickness variations will typically not influence the plasmon-enhanced colorimetric histology image. Ultramicrotome is capable of producing sections of incredible uniformity e.g. root-means-square (rms) variations (according to manufacturer specifications) across whole slide sections of 2-5 %. Here, the thickness uniformity of tissues generated using the ultramicrotome was experimentally confirmed both via plasmon enhanced colorimetric histology of the homogeneous resin and Atomic Force Microscopy (AFM). A series of AFM measurements conducted on the nominally (according to the microtome settings) 100 nm thick sections revealed a thickness variation of $95.7 \pm 4.8$ nm (s.e.m) on the plasmonic devices (see Fig. S2).

**Fig. 3: Plasmon-enhanced array tomogram of polystyrene microsphere test sample.**

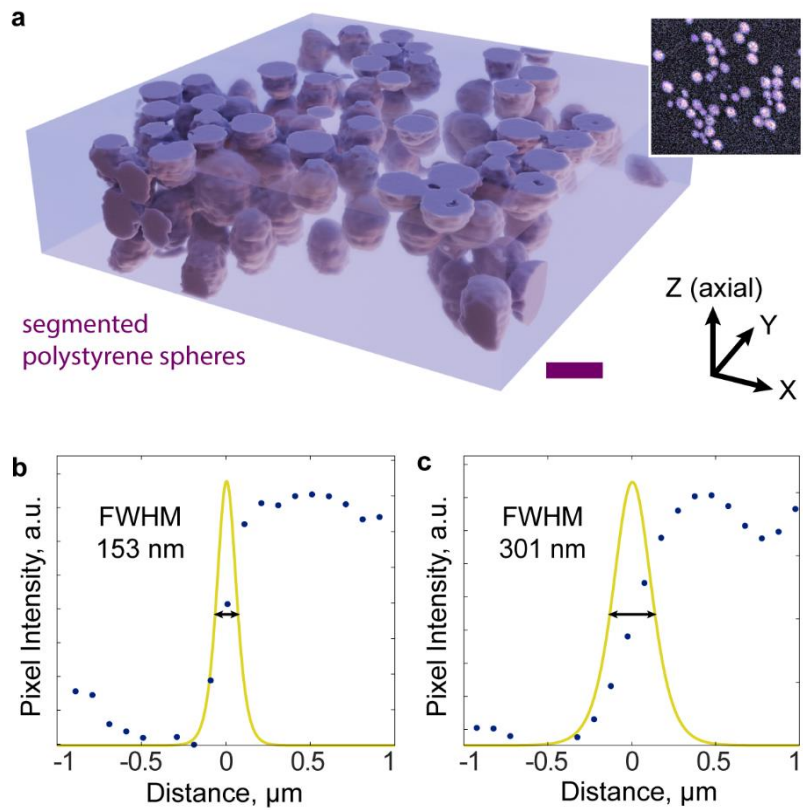

a) Optical PE array tomography volume render of resin-embedded polystyrene microspheres. The top right corner shows a section of the spheres. Scalebar is 5 μm. b) Resolution analysis in the axial (Z) and c) lateral direction. An example profile of a representative sphere edge is plotted in blue and the corresponding derivative of the fitted sigmoid function used to estimate spatial resolution is plotted in yellow. The double headed arrow represents the Full Width at Half Maximum (FWHM).

Following the demonstration of PE array tomography using our model sample, the technique was then applied to imaging of a 'real' histological sample consisting of human malaria-infected Red Blood Cells (RBCs). Some of the infected cells contain hemozoin crystals which can result from the malaria parasite digesting hemoglobin within the host cell. These samples were characterised using plasmon-enhanced colorimetric histology (Fig. 4a) and array tomography (Fig. 4b) as well as SEM imaging (Fig. 4c). Due to the conducting surface of the plasmonic devices no treatment of the sample (e.g. deposition of a conducting layer) was necessary in order to generate electron microscopy images. This enabled optical plasmon-enhanced imaging and electron microscopy to be performed and registered on the exact same histological sample (Fig. 4d). The ultramicrotome section thickness for the RBCs was maintained at 100 nm. We observe many of the features are correlated between the plasmon-enhanced colorimetric histology and SEM, however, the PE array tomography provides a complimentary color contrast mechanism which aids in differentiating specific chemically distinct features in 3D, such as healthy and malaria infected RBCs as well as the hemozoin crystals in the infected RBCs.

**Fig. 4: Plasmon-enhanced array and electron array tomography of RBCs.**

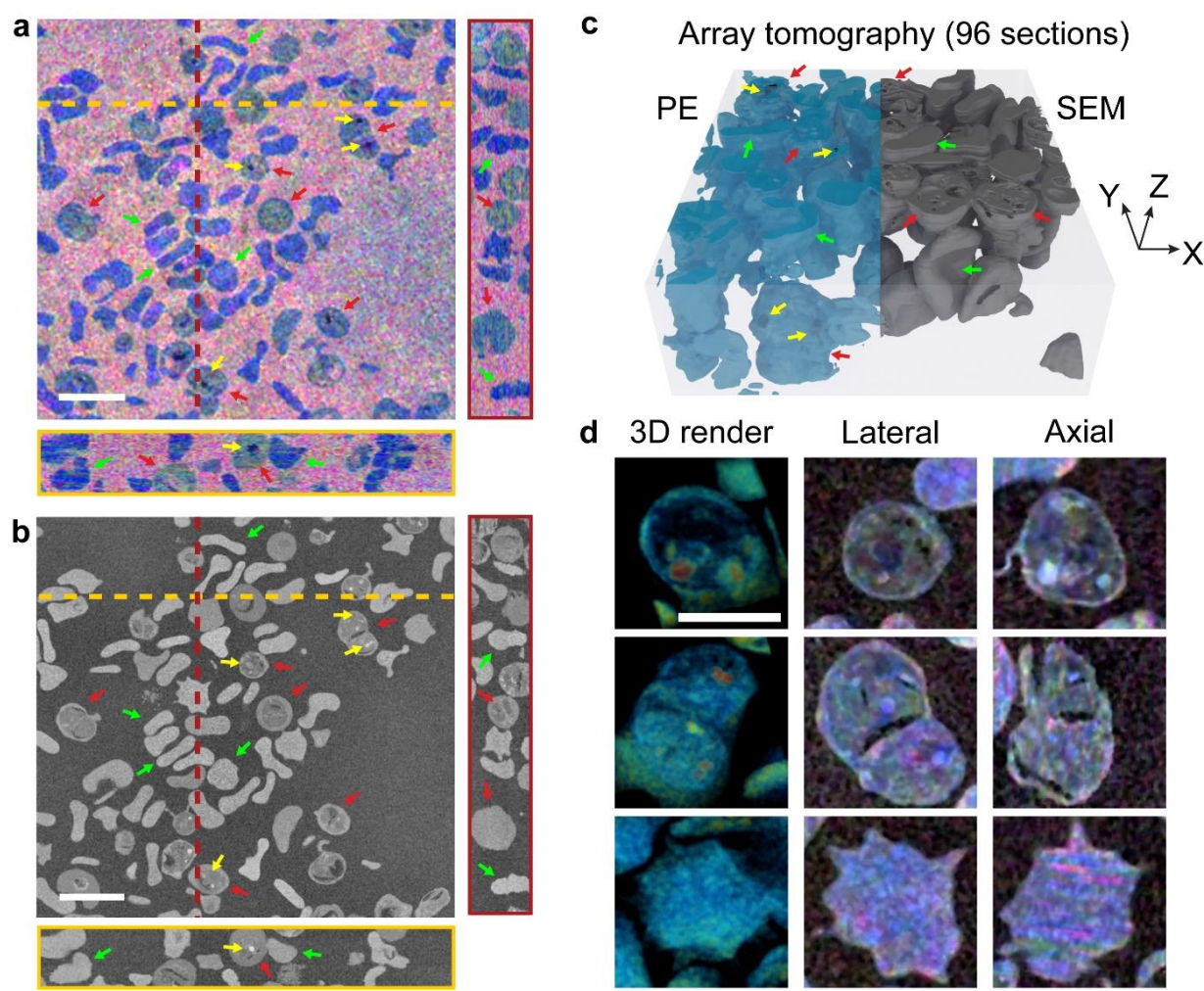

Malaria infected red blood cells as imaged with array tomography (96 serial resin sections, each 100 nm thick). a) Optical plasmon-enhanced colorimetric histology example XY section and XZ and YZ cross sections with their positions denoted by red dashed lines. b) Correlative SEM results for the same sample. c)3D isosurface render of the reconstructed red blood cells . Yellow arrows in (a), (b) and (c) point to the location of the hemozoin crystals, green and red arrows indicate healthy and malaria infected RBC respectively. d) Combination of PE array tomography and electron array tomography for three isolated cells. First column shows a 3D (false color) render. The second and third column show the combined images in both lateral (XY) and axial directions (XZ/YZ). The color contrast shown is produced directly from the plasmonic effect whilst the image brightness is from SEM. Scalebars in (a, b) are 10 um long. Scalebar in (d) is 5 um.

Malaria infected RBCs appear of iregular round shape, torquise in color with dark blue hemozoin crystals inside, whereas the healthy ones have the characteristic disc-like shape distiguished by the blue color. In order to further illustrate the additional complementary information that plasmon-enhanced imaging can provide to electron microscopy, spectral data from the hemozoin crystals and RBCs was collected. From these data, we generated a PE array tomography rendering (Fig. 5a) where we observed significant differences between the spectrum collected from the hemozoin crystals, RBCs, and surrounding resin matrix (Fig. 5b). These spectral differences between intracellular components manifest as a colorimetric differential when viewed under the optical microscope (Fig. 5c).

**Fig. 5: Spectra of hemozoin crystals and surrounding RBC.**

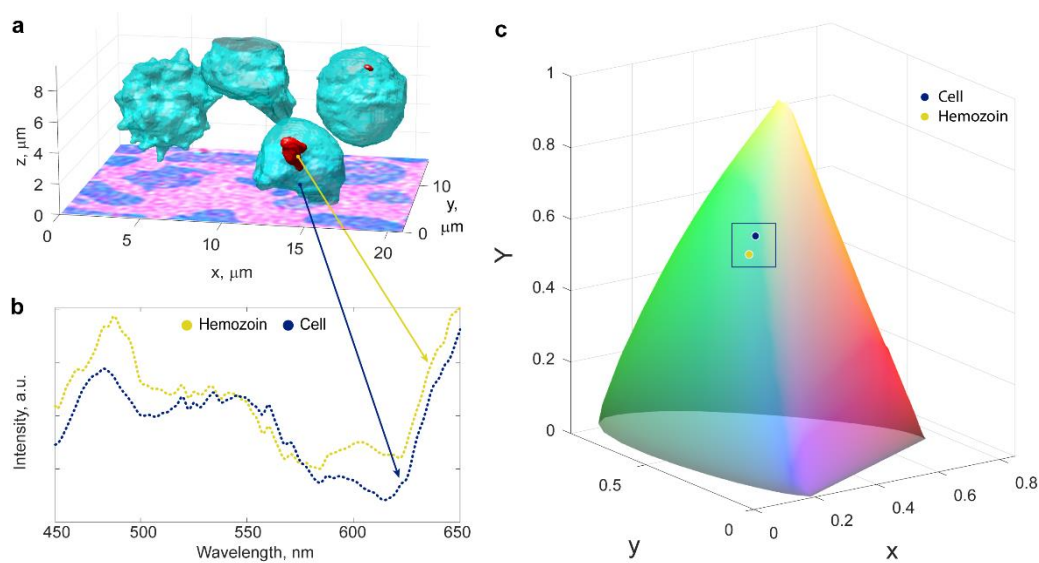

a) Surface rendering of the RBCs (turquoise) and hemozoin crystals (red) obtained via optical PE array tomography. XY image is a raw image of an example section. b) plots of spectra extracted from the RBC (blue) and hemozoin crystal (yellow). c) Corresponding color outputs plotted in the CIE 1976 xyY color space on a 3D chromaticity diagram, with x and y specifying the chromaticity and Y the luminescence. The dark blue circle represents an average color for the RBC, the red-filled circle represents the color observed for the hemozoin crystals.

The same technique was also applied to imaging tissue sections from a mouse in which a disease mimicking the central nervous system (CNS) autoimmune disorder multiple sclerosis had been induced. Sampling was performed from an optic nerve (ON) at disease onset, namely 10 to 12 days post disease induction (Fig 6).

**Fig. 6: Plasmon-enhanced array and electron array tomography of optic nerve.**

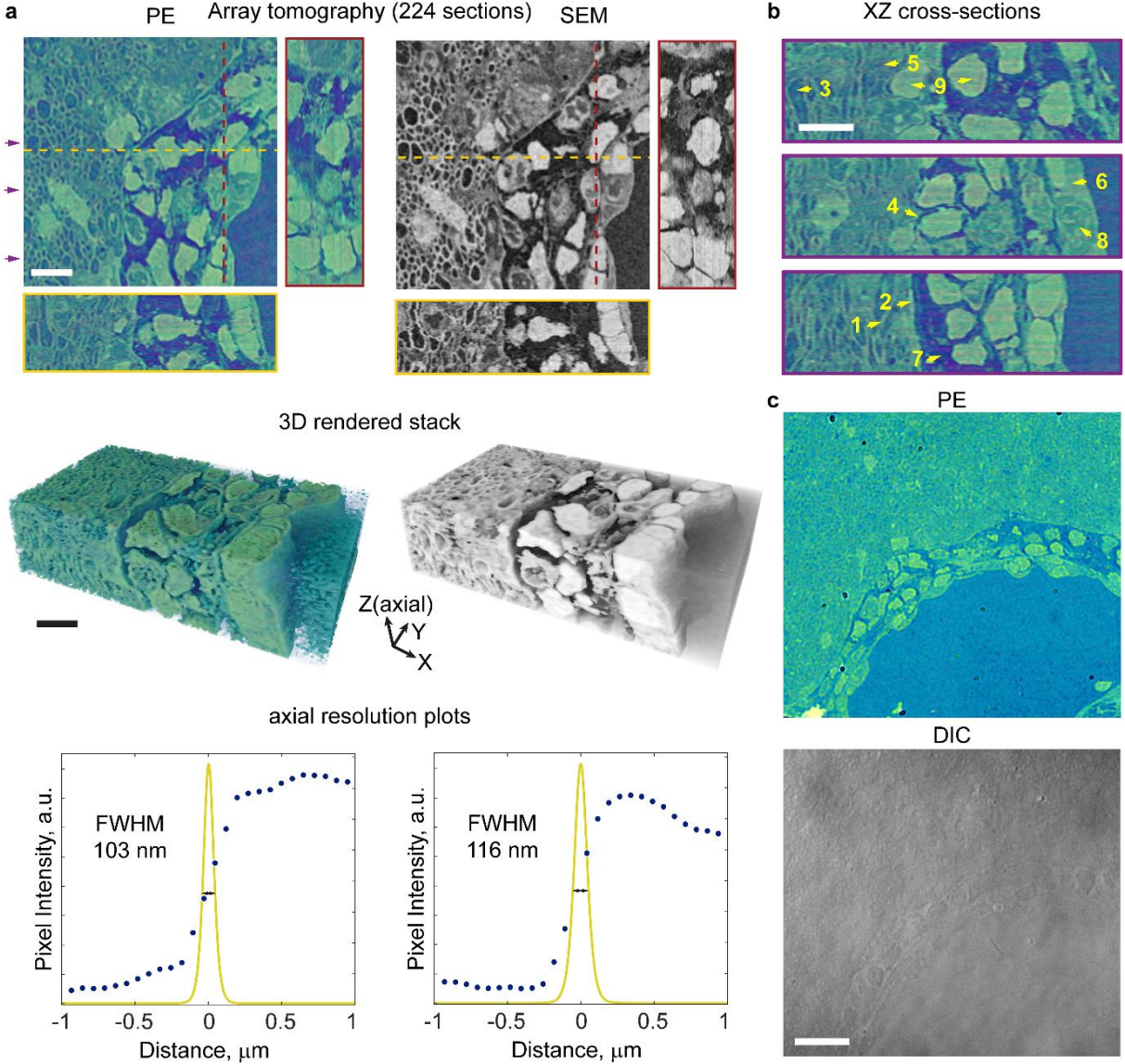

Optic nerve from a mouse with a CNS autoimmune disorder multiple sclerosis (224 serial resin sections, each 75 nm thick). a) Optical plasmon-enhanced colorimetric histology and correlative SEM results of a representative XY section, and XZ and YZ cross sections with their positions denoted by yellow and red dashed lines respectively. Bellow their respective 3D isosurface renders and the axial resolution plots. b) XZ cross sections from the stack taken at the positions denoted by arrows in (a). (1) myelinated axon, (2) glia limitans, (3) macrophage, (4) lymphocyte disrupting and entering the glia limitans, (5) granulocyte, (6) lymphocyte interacting with endothelial cells, (7) perivascular space, (8) endothelial cell, (9) lymphocyte. c) PE and DIC microscopy image of a 75 nm thick unstained nerve tissue section on plasmonic device and conventional glass microscope slide respectively. Scalebar in (a, b) is 10 um. Scalebar in (c) is 30 um.

Similarly to the RBC sample, the ON samples were characterised using plasmon-enhanced colorimetric histology (Fig. 5a) and array tomography as well as SEM imaging performed on the same histological sample with the ultramicrotome section thickness set at 75 nm. Notably, numerous features exhibit a correlation between plasmon-enhanced colorimetric histology and scanning electron microscopy (SEM). However, the plasmonic-excitation (PE) array tomography introduces a complementary color contrast mechanism, facilitating the differentiation of specific chemically distinct features in (3D) associated with different processes, such as the interaction of the immune system cells with the ON constituents. At 10 to 12 days post disease induction, cells of the immune system (or inflammatory cells) targeting the CNS are already evident, but in relatively moderate numbers with minimal infiltration of brain and ON tissue. Therefore, multiple features of normal CNS tissue can be identified (Fig 6b). This includes cerebral endothelial cells (8), which are specialized cells lining the blood vessels, the peri-vascular space (7) and the glia limitans (2), a layer which forms the outer edge of the CNS tissue. Within the ON tissue, glial cells (any cell other than nerve cells, Fig 3S), characterized by their dark cytoplasmic, can be observed as well as axons ensheathed within myelin (1), which vary considerably in terms of their diameter and density across given tissue section (Fig 3S). This difference is readily observable in the PE imaging, while completely missed in the SEM imaging. The inflammatory cells are found to be associated with endothelial cells (8), or to have traversed into the perivascular space of the large blood vessel (4-5). Inflammatory cells consist of multiple sub-types, distinguishable by morphological features as well as staining intensity. For example, a single granulocyte characterized by its multilobular nucleus (5), macrophages (3), characterized by their large size and large nuclei and lymphocytes (9), characterized by their round shape and dense staining are clearly distinguishable.

## 3. Discussion

The overarching goal of this work was to demonstrate plasmon-enhanced array tomography in which super-resolution axial imaging can be achieved through the application of tissue ultramicrotomy. The contrast enhancement provided by plasmon-enhanced colorimetric histology enables label-free high-contrast images to be generated of histological tissue sections that are much thinner than the diffraction limit. To demonstrate this, we have imaged ultrafine sections of resin-embedded cells and tissues in the range of 30 to 200 nm thickness. Unlike current phase imaging techniques, our approach does not rely on any interference effects or sample birefringence, and in our tests, PE colorimetric histology was able to dramatically outperform the image contrast of competing phase-contrast methods such as DIC for ultrafine (< 200 nm) histological sections (e.g. Figs 2b and 6c).

Although the use of serial ultrathin tissue sections naturally lends itself to axial super-resolution, as has been demonstrated in EM tomography, our well-characterised test sample composed of microspheres was used to experimentally confirm an optical axial resolution of 153 nm which is around 1.5 times the tissue section thickness, and 103 nm for 75 nm tissue sections. This can be further improved for sections down to 30 nm in thickness (Fig 1S). The conducting nature of the plasmonic slides allowed for correlative EM imaging of human RBCs where intracellular features were resolved, including the hemozoin crystals. Whilst the intracellular morphology of the sample is visible using high-resolution EM, the grey level contrast limits the available dynamic range making segmentation challenging. The benefits of utilising PE induced color contrast, which can be obtained using a standard optical microscope, are confirmed when looking at the spectral characteristics of the cell which underlie these colorimetric differences. Similar results were obtained from large tissue sections (e.g. optic nerve), where biological processes can be assesed with ease on very large section (>1 mm) with 103 nm axial resolution.

Another significant benefit of PE array tomography is that it is entirely compatible with other standard super-resolution optical imaging techniques such as confocal microscopy or structured illumination microscopy (SIM). This means that extension of this technique to label-free super-resolution microscopy in all three dimensions is entirely feasible. This would allow super-resolution label-free imaging in both the lateral and axial directions over all relevant intracellular lengthscales which are typically beyond the ability of conventional brightfield microscopy to resolve. The potential applications for optical PE array tomography are vast ranging from improvements in cancer diagnsosis[32] to early detection of neurological diseases[33].

## Methods

*Plasmonic slide fabrication*

The plasmonic slides were fabricated at the Melbourne Centre for Nanofabrication (MCN). The substrate was cleaned borosilicate glass wafers. A 6 nm layer of chromium was deposited onto the glass via electron beam evaporation. A 100 nm layer of silver was subsequently deposited by the same method. The nanoscale circular apertures were patterned into the films via displacement Talbot lithography ono a square array with a resolution of 150 nm. The lateral aperture periodicities were 400 nm in both orthogonal directions.

*Optic nerve, RBCs and polystyrene sphere sample preparation*

The optic nerve samples were fixed using 2.5 % (v/v) glutaraldehyde with osmium tetroxide as a post-fixative for the mouse tissue. Serial dehydration was performed using ethanol and acetone before embedding the tissue in epoxy resin. The polystyrene microsphere test sample was prepared using crosslinked microspheres purchased from Cospheric with diameters of 4.8-5.8 μm in Spurr resin which was cured at a temperature of 60°C over three days. The red blood cell sample comprised of a mix of healthy and malaria infected cells was prepared for SEM imaging via reduced osmium fixation comprising 0.1 M sodium cacodylate containing 5 mM calcium chloride [34]. Glutaraldehyde (2.5% v/v) was used to fix the cells, which were then rinsed in 0.175 M sodium cacodylate. Post-fixation used potassium ferricyanide reduced osmium tetroxide followed by rinsing with distilled water, then treatment with 1% (X/v) thiocarbhohydrazise. Further osmication was performed using 2% (X/v) non-reduced osmium tetroxide followed by en-bloc staining with 1% (X/v) uranyl acetate before dehydration and embedding in epoxy resin.

*Ultramicrotome sectioning*

Sectioning was performed using a Leica EM UC7 ultramicrotome. Samples were first trimmed coarsely using a razor blade to expose the specimen within the resin block followed by trimming to a small (~15 × 15 mm) rectangle using a 25 mm glass microtome knife. A Diatome ultra 45 jumbo diamond knife filled with MilliQ water was used for sectioning at a cutting speed of 1 mms$^{-1}$ and a cutting angle of zero degrees. Sections between 30 nm (see Fig. S1) and 200 nm were produced in ribbons along the knife edge. These were later detached and manoeuvred to the water edge on the plasmonic slide using an eyelash brush before draining the water from the boat. The plasmonic slide was then removed from the knife boat and baked on a hotplate at 120 °C for 15-30 minutes to dry and remove any excess water.

*Optical characterisation*

Optical imaging was performed using a Nikon Eclipse Ti-2 E optical microscope equipped with a DS-Ri2 camera. Both 40× and 100× objective lenses with NA values of 0.65 and 0.95, respectively were used for image capture. White levels were set by defining a region of the plasmonic slide with no sample in place. DIC images were captured Zeiss axio observer optical microscope equipped with an axio cam CCD camera. Samples were not coverslipped. Optical spectra were captured using an NKT photonics SuperK COMPACT white light laser source paired with a SuperK VARIA tunable filter. Data was collected using a 5 nm bandwidth in 5 nm steps over the range of 450 nm to 650 nm.

*Image processing*

Images were processed using a combination of the Fiji distribution of ImageJ 2.1.051, the microscopy image browser (MIB) MATLAB-based software package, and MATLAB 2022a. ImageJ was used for stacking and aligning images. In order to display the volumetric data, it was necessary to laterally interpolate the optical and SEM data by approximately a factor of three such that the sampling in the X-Y plane matched the Y-Z and X-Z planes. Histogram equalisation was implemented in MIB. Alignment of the image stack was achieved using the linear stack alignment with SIFT plugin available in Fiji. Drift correction to ensure alignment of array slices was performed in MIB to refine the output of the SIFT algorithm.

Spatial resolution analysis performed in MATLAB involved calculating the derivative of a sigmoid function fitted to lineouts that were taken across the edge of the ultramicrotome sections (since this provided a nominally sharp edge). The FWHM of Gaussian's fitted to this derivative was used to provide an experimental estimate of the lateral and axial resolution. The values quoted are the average of six separate resolution measurements, and errors are the standard deviation.

For volumetric rendering, a 3D median filter (medfilt3()) with $\sigma = 3$ px was applied in MATLAB, followed by visualisation using the Vol3d function. Correlative renders were produced using OmooLab's BioxelNodes add-on for Blender 4.2.

Optical transmission spectra were normalised to correct for any differences in exposure time and divided by spectra collected from the bare plasmonic slide. For the spectra displayed in the paper, data from 100 pixels were averaged in order to improve the statistics.

Edge visibility for lineouts was calculated as the difference of the max and min values divided by the sum of max and minimum values near the edge.

$$Edge\ Visibility = \frac{max - min}{\max + \min} \tag{1}$$

*Alignment of image stacks*

We used a deep learning-based 3D nanoaperture radiance field (NRF) framework to optimize captured image stacks via simultaneous color correction, denoising, and alignment. The NRF employs an end-to-end gradient descent optimization to integrate these tasks within a cohesive pipeline, enhancing workflow efficiency and output quality.

Building on neural radiance field advancements, the NRF framework represents image stacks as a continuous 3D model, preserving spatial and color information. To simulate in-camera image signal processing (ISP), a camera embedding mechanism accounts for device-specific factors affecting color, enhancing the realism and consistency of reconstructed images. A learnable affine transformation per image is also included, optimizing rotation, scaling, and translation parameters to precisely align images spatially.

Following NRF optimization, we extracted final images using a tailored post-processing approach, addressing detail limitations inherent in direct rendering. Specifically, we computed per-channel color transformations between rendered and captured images, applying these transformations to correct the captured images' color accurately. Spatial alignment was maintained by applying the optimized affine transformations. To reduce noise, regions identified by comparing rendered and corrected images had their noisy pixels replaced by corresponding rendered pixels, ensuring structural integrity and producing high-quality, aligned, and color-accurate outputs.

*Atomic Force Microscopy characterisation*

Atomic Force Microscopy (AFM) measurements were performed using an Asylum research MFP3D-SA AFM with Nanosensors pointprobeplus cantilevers in non-contact mode. No additional sample preparation was required. A scan rate of 1 Hz was used for scan areas smaller than 5 μm × 5 μm while a scan rate of 0.3 Hz was used for scan areas larger than 5 μm × 5 μm. The set point was fixed at 640 mV with an integral gain of 8.5. A drive amplitude of 220 mV was used for the section surface and increased up to 720 mV when scanning tissue section edges. AFM image processing was performed in Gwyddion 2.59 in which backgrounds were leveled by mean plane subtraction for those scans that did not feature any section edges and leveled by fitting a plane through three points for scans that did feature section edges. Horizontal scar correction was applied to all images.

*Scanning Electron Microscopy (SEM) characterisation*

Plasmonic slides were mounted to flat SEM sample holders using a small piece of conductive carbon tape on the glass side of the slide with an edge of the tape folded to make contact with the silver side. Slides were cleaned with compressed nitrogen prior to entering the SEM vacuum chamber. SEM imaging was performed using a Hitachi SU7000 ultra-High-Resolution Schottky Scanning Electron Microscope.

The red blood cell sample was imaged with a beam voltage of 1 keV at 900x magnification for a resultant pixel size of 27.6 nm. The optic nerve sample used a beam voltage of 0.5 keV and magnification to 150× yielding a 82.7 nm pixel size. The instruments 'middle detector', a type of back-scattered electron detector, was used as it provided the best tissue contrast for the selected beam voltages.

**Acknowledgements**

acknowledge the support of the Australian Research Council Discovery Project (DP220103679).

**Data Availability**

All the other data used in this study are available in the article and its supplementary information files and from the corresponding author upon request.

**Supplementary Information**

Supplementary Images

**Fig. S1: Comparison of optical images collected from 30 nm thick sections from polystyrene microsphere test sample on glass and plasmonic substrates.**

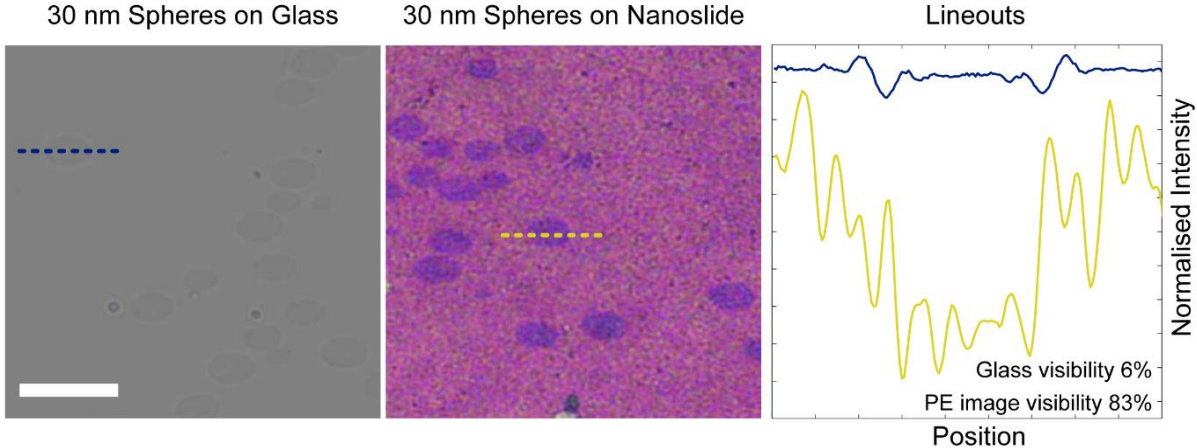

Normalised lineouts obtained from 30 nm section of resin-embedded polystyrene microspheres section imaged using standard brightfield microscopy on glass (left) and plasmonic (middle) substrates. Some edge contrast (light and dark fringes) for the microspheres can be faintly discerned in the lineouts (right) due to the differences in RI between the microsphere edge and the resin. An important distinction is that on glass inside the spheres, the intensity is similar to the substrate indicating only the sharp edge of the sphere is discernible, whereas on the plasmonic the intensity remains distinct from that of the surrounding resin. The relative visibility values of the brightfield images and PE image are shown in the bottom right-hand corner. The white scale bar is 5 μm.

**Fig. S2: Example Atomic Force Microscopy (AFM) analysis of tissue section thickness.**

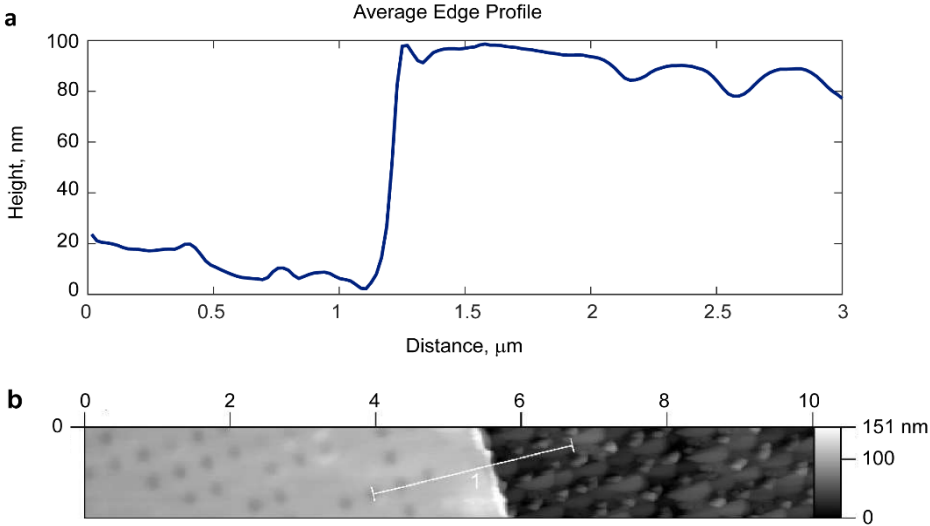

AFM scans of nominally 100 nm thick optic nerve sections. The example result shown was produced by averaging 10 independent AFM line scans from different regions. a) and b) show the tissue section thickness and tissue sample surface, respectively, on a plasmonic substrate.

**Fig. S3: Plasmon-enhanced and electron imaging of whole optic nerve section.**

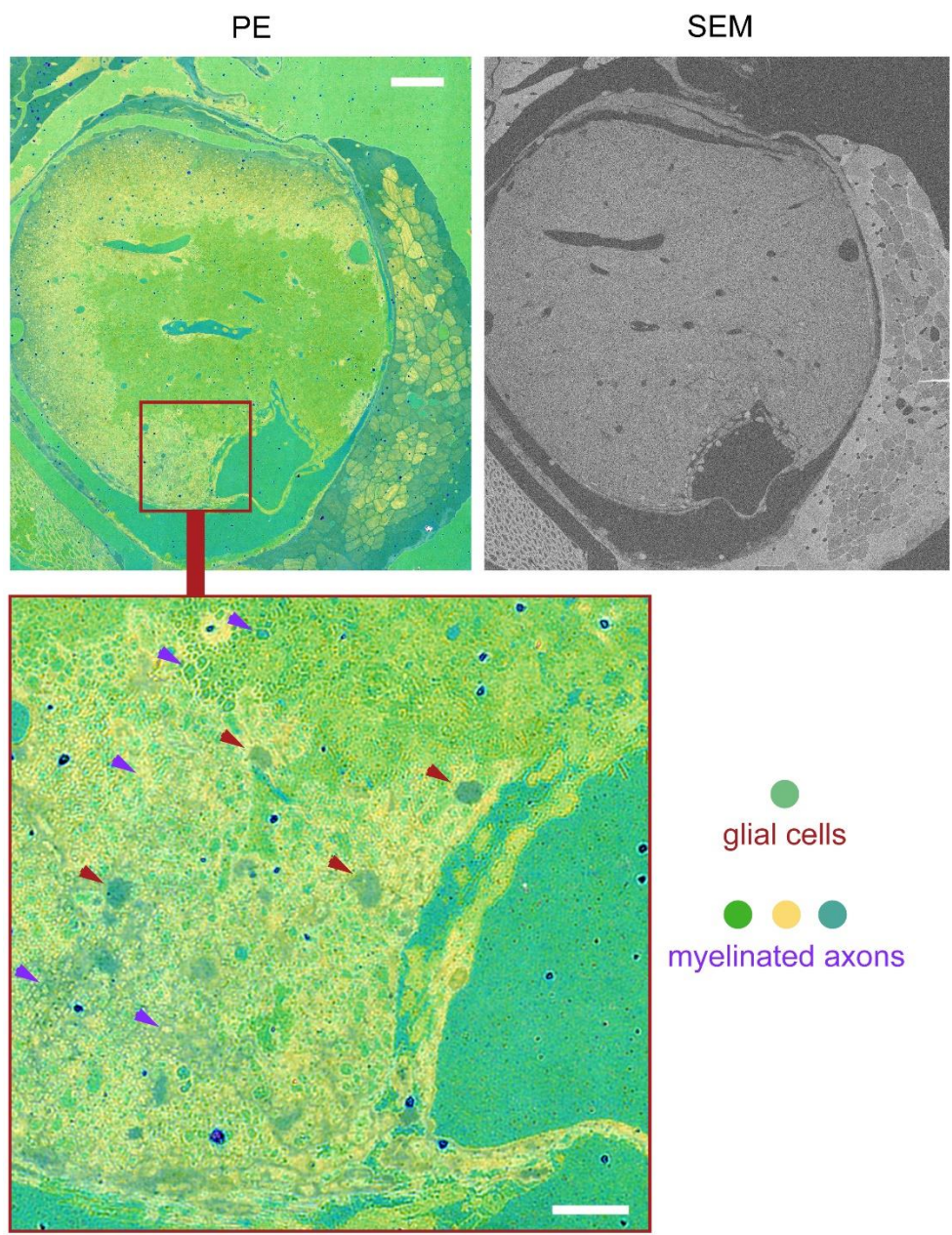

Optic nerve from a mouse with a CNS autoimmune disorder multiple sclerosis 100 nm thick. Optical plasmon-enhanced colorimetric histology and correlative SEM results of a representative XY section. The arrows show glial cells and myelinated axons with different densities and size. The scalebars are 100 and 20 μm respectively.